# Some Aspects of Electron Dynamics in Electron Cyclotron Resonance Ion Sources

V. Mironov, S. Bogomolov, A. Bondarchenko, A. Efremov, and V. Loginov

*Joint Institute for Nuclear Research, Flerov Laboratory of Nuclear Reactions, Dubna, Moscow Reg. 141980, Russia*

Abstract

*Electron dynamics in Electron Cyclotron Resonance Ion Source is numerically simulated by using Particle-In-Cell code combined with simulations of the ion dynamics. Mean electron energies are found to be around 70 keV close to values that are derived from spectra of X-ray emission out of the source. Electron life time is defined by losses of low-energy electrons created in ionizing collisions; the losses are regulated by electron heating rate, which depends on magnitude of the microwave electric field. Changes in ion confinement with variations in the microwave electric field and gas flow are simulated. Influence of electron dynamics on the afterglow and two-frequency heating effects is discussed.*

## I. Introduction

Electron Cyclotron Resonance Ion Source (ECRIS) [1] is a minimum-B magnetic plasma trap. Plasma in the source is heated by microwaves absorbed by electrons in vicinity of the ECR surface. Such sources are mainly used to produce intense beams of highly charged ions, which require relatively high mean electron energies (in the keV range) and long electron (ion) confinement times (~ 1 ms) in the plasma. To understand and to optimize the source performance, it is crucial to know main parameters of the plasma electron component, such as the electron energy distribution function (EEDF) and the spatial distribution of electron density.

Experimental studies of the electron component in ECRIS are mainly focused on detection of X-rays emitted when electrons collide with plasma ions or are lost at plasma chamber walls [2,3]. Measurements with Langmuir probes [4] are only possible at peripheral parts of the plasma, not in the dense plasma inside the region limited by the ECR surface. Substantial amounts of relatively cold electrons with mean energies of around 10 eV are detected with the probes, corresponding to the positive plasma potential of 20-50 V.

The so-called spectral temperature ($T_{eh}$) is extracted from X-ray spectra; typical values are in the range of a few tens of keV [5]. The temperature strongly depends on the source magnetic field configuration and on frequency of microwaves. No direct correlation is seen between $T_{eh}$ and extracted currents of ions: when injected microwave power is varied by an order of magnitude with the corresponding changes in the extracted currents, the spectral temperature remains constant or varies slowly [3]. Also, reaction of $T_{eh}$ on gas flow into the source is weak. Intensity of X-rays varies strongly, with higher count rates for higher injected microwave power and gas flow in the source. Moreover, decay times of the emission intensity and $T_{eh}$ after switching the microwave heating off are long, ~(10-100) ms, which is difficult to match to the estimated life times of ions inside the ECRIS plasma [2, 6].

Shirkov describes the electron component in ECRIS [7] as consisting of three fractions: cold electrons with energies below 100-300 eV, warm electrons with keV energies and hot electrons with energies that correspond to the spectral temperature. We change here the original terminology of [7] to follow more popular designations. The cold electrons are directly produced as secondaries in ionizing collisions and effectively produce the lowly charged ions, the warm electrons constitute main body of the electron component and are responsible for creation of the highly charged ions in ECRIS, the hot electrons stabilize plasma due to their good confinement and can be considered as a tail of the warm electron EEDF. This fractionalization is often used when analyzing ECRIS operation.

Several groups performed numerical simulations of the electrons in ECRIS by using Fokker-Planck (FP) or Monte-Carlo Collisions Particle-in-Cell (MCC-PIC) codes. Neri et al. [8] use their MCC-PIC code to simulate 3D plasma dynamics and obtain EEDF that generally agrees with the Shirkov's description, with warm electrons being 80% of the total electron population. In the code, electrons are followed for 10 μs only, and collisional losses of electrons are simulated as a single collision scattering that rotates electron trajectory by 90°.

Cluggish et al. [9] studied with 1D bounce-averaged FP code the plasma response to variations in gas pressure and microwave power. They report that the mean electron energy can be as high as ~40 keV at relatively small microwave powers (~100 W).

Heinen et al. [10] followed the collisionless electron movement and heating in ECRIS. Electrons were traced for 1 μs only, and the calculated energy spectra are close to what is suggested in [7]. The simulations of Biri et al. [11] with their TrapCAD code give essentially the same results. Strong localization of electrons within the ECR volume is seen after electron heating, as well as electron energy gradients that depend on type of electron trajectories in respect to the resonance surface.

So far, our NAM-ECRIS model [12,13] was applied for simulations of ECRIS in assumption that all electrons are *warm*. The electron temperature $T_{ew}$ was taken as a free parameter for the code, being connected to the coupled microwave power by assumption that the mean energy of electrons that are lost out of the plasma is $3/2T_{ew}$. The electron life times were calculated by using the analytical estimation of Post [14], which is a strong simplification. More detailed information on electrons in ECRIS is needed to improve accuracy of the model predictions. For this purpose, we develop the special supplementary code to calculate the electron component dynamics. Space-averaged results of the code are the mean electron energies and confinement times. These results are then used as input for those parts of NAM-ECRIS that calculate the ion dynamics in ECRIS.

## II. Model

Our Particle-in-Cell code traces dynamics of moderately large number of numerical particles ($10^3$) that represent electrons. Particles move in the magnetic field calculated for DECRIS-SC2 18 GHz source [15]. The magnetic field profile is optimal for production of $Ar^{8+}$ ions; experimentally, up to 1 mA of $Ar^{8+}$ current had been extracted from the source. The inner diameter of the source chamber is 7.4 cm, the chamber length between the biased disk (Ø2 cm) and extraction electrode (Ø1 cm) is 28 cm. Hexapole magnetic field at the radial wall is 1.1 T, magnetic fields at the axis at the injection and extraction sides of the chamber are 1.97 and 1.35 T respectively, the minimum field is 0.47 T.

Before starting the electron simulations, the ion part of NAM-ECRIS is run to simulate argon ECRIS plasma. Inputs for the ion simulations are (1) the electron temperature inside the relativistically broadened ECR zone $T_{ew}$, (2) numerical weight of particles, which defines the gas flow into the source, and (3) potential dip, which retards ions inside the ECR zone. Outputs of the code are extracted ion currents, globally-defined ion life time, the ion and electron densities calculated on rectangular numerical mesh, and array of initial positions and energies for the secondary electrons created in ionizing events. Energies of electrons are taken to be equal to the ionizing potentials of the atom or ion that was ionized in the corresponding event [16]. The assumed localization of energetic electrons inside the ECR zone means that most of the secondary electrons are created inside the zone being distributed almost isotropically there.

The electron temperature in NAM-ECRIS(i) is taken as an input in iterative way basing on the results of NAM-ECRIS(e) simulations. Typical output parameters of ionic part (input parameters for the electronic part) correspond to the extracted currents of $Ar^{8+}$ ions of around 1 mA, electron density ~$10^{12}$ cm$^{-3}$, mean charge state of ions inside the dense plasma ~4, mean $n_i Z^2$ ~$10^{13}$ cm$^{-3}$ and mean ion life time ~0.5 ms.

Electrons in the electronic part of the code are launched according to the prepared array of initial positions, their velocities are chosen according to the primary energies with random orientation of the velocity vectors.

The Boris mover is used to calculate particle motion in the magnetic field [17]. Electrons bounce along the magnetic field lines and are trapped inside the source by mirror force if their velocities are outside the loss cone in the velocity space. Curvature drift causes the electron movement across the magnetic field lines. When bouncing, the particles periodically cross the ECR zone, where gyro-frequency of electrons equals to the microwave frequency, which corresponds to the magnetic field $B_{res}$=0.643γ [T] (γ is the Lorentz factor) for 18-GHz microwaves.

Whenever the particle crosses the ECR zone, it experiences random kick in the direction perpendicular to the magnetic field line. The kick value V is calculated according to Lieberman and Lichtenberg [18] as

$$V = \frac{e}{m_e \gamma} E_0 t_e \cos(x) \quad (1)$$

Here, $E_0$ is the magnitude of the applied microwave field at the resonance field, x is a random variable in the range from 0 to 2π giving the phase between the velocity vector and the electric field, e and $m_e$ are the electron charge and mass, and $t_e$ is the effective time the particle spends in resonance. For the time we select the minimal value of two different times, $t_e = min(t_{e1}, t_{e2})$:

$$t_{e1} \approx 1.13 \left(\frac{2}{|\alpha v_\parallel| \omega}\right)^{1/2}; t_{e2} = (0.71/\omega)(2\omega/\alpha v_\perp)^{2/3} \quad (2)$$

Here, ω is the microwave angular frequency, $\alpha = B_s^{-1}(dB_s/ds)$ is the normalized magnetic field gradient along the magnetic field line, $v_\parallel$ and $v_\perp$ are the velocity components along and perpendicular to the magnetic field line respectively. The first value corresponds to the case when particle passes with constant axial velocity through the resonance zone, while the second expression is applied when the particle begins to turn in the resonance zone.

Magnitude of the microwave electric field is a free parameter in our calculations. In the real conditions, the field depends on the power of the injected microwaves ($E_0$~$P^{1/2}$), geometry of chamber, density of the plasma inside the source and other factors. Electric field of microwaves has a complicated spatial dependence of magnitude and phase [19]. We omit at the moment all these details and consider the field constant over the ECR surface.

Random kicks of electrons at ECR result in diffusion in the velocity space and in heating of the electron component. It is important to note that the heating rate is slowing down with increasing the electron velocity because the time interval for electrons to resonate with microwaves when passing through ECR is decreasing.

Each computational time step the particles are scattered by a small angle in random direction due to electron-ion and electron-electron collisions. The angle of scattering Θ for colliding species α and β is selected randomly according to Takizuka and Abe's collision method [20] through a Gaussian random variable δ related to Θ by

$$\delta = \tan(\Theta/2) \quad (3)$$

where δ has zero mean value and the following variance:

$$\langle \delta^2 \rangle = \left(e_\alpha^2 e_\beta^2 n_L \log \Lambda \big/ 8\pi\varepsilon_0^2 m_{\alpha\beta}^2 u^3\right)\Delta t \quad (4)$$

Here, $e_\alpha$ and $e_\beta$ are electric charges for the species α and β, $n_L$ is the smaller density of the particle species α and β, logΛ is the Coulomb logarithm, $u = |v_\alpha - v_\beta|$ is the relative speed, Δt is the time step, and $m_{\alpha\beta}$ is the reduced mass: $m_{\alpha\beta} = m_\alpha m_\beta / (m_\alpha + m_\beta)$.

For the electron-ion collisions ($m_{\alpha\beta} \approx m_e$ and $u \approx v_e$) the deflection angle can be calculated as a single event proportional to the sum of individual contributions of scattering on ions with density $n_{iQ}$ and charge state Q, such that the variance of δ-factor is:

$$\langle \delta^2 \rangle \sim \frac{\sum_Q n_{iQ} Q^2}{v_e^3} \quad (5)$$

The deflection angle is decreasing fast with increasing the electron velocities.

For the electron-electron collisions, the angle is proportional to the electron density in the corresponding computational cell. To handle these collisions, colliding particles are paired within the cell according to the algorithm of Takizuka-Abe. This procedure is time consuming; we neglect the electron-electron collisions in most cases taking into consideration that the electron-ion collision frequency is much higher than for the electron-electron collisions for argon plasma in typical conditions.

Whenever the computational particles leave the source chamber, its velocity component along the magnetic field line is calculated. If the corresponding energy is less than 25 eV for collisions with the walls, 25 keV for particles hitting the extraction aperture, or 250 V for particles that hit the biased electrode at the injection side of the source, the velocity component changes its sign and the particle is reflected back into the source. In this way we take into account the electron retardation either by the positive plasma potential or by potentials at the extraction aperture and biased electrode. If electron is lost, it is returned back by placing into point that is randomly selected from the array of initial positions, with the corresponding initial energy. Energy of the lost electron is saved as well as time of the event. Simultaneously, we randomly select one particle in the computational domain and decrement it's energy by energy of the newly created electron multiplied by factor of two. In this way, energy losses due to ionization processes are taken into account.

From flux of the lost particles, the mean electron life time is calculated, as well as the mean energy of the lost electrons. Also, we calculate the mean energy of the electrons that are inside the plasma.

Typical calculation runs up to (0.1-5.0) ms of physical time depending on the requested statistics. The computational time of one run is up to 24 hours by using Intel® Core™ i3-4340 @3.6 GHz CPU.

### III.     Results

We begin with presenting the results obtained in assumption that ion density and mean charge state are constant everywhere in the plasma, specifically $n_iQ^2=10^{13}$ cm$^{-3}$. In Fig.1, there are shown the first one hundred microseconds of the time evolution for the electron losses (left scale) and for the mean electron energy (right scale). A word of caution is that these curves should not be considered as a simulation of plasma ignition, but they are used only to show the relative importance of the processes responsible for the electron losses and heating in ECRIS. In the beginning, we do not apply heating processes and microwave electric field is set to zero. Two distinct situations are presented – with and without the electron scattering in collisions with ions, black and red curves correspondingly.

Initial distribution of electron velocities contains a large fraction of particles within the loss cone. These particles leave the system almost immediately, which results in the sharp peak in electron loss rate soon after starting the simulations. The lost electrons are returned back and there is a certain possibility that new born electron is outside the loss cone, so the loss rate is decreasing. For the case with electron scattering, the loss rate saturates at some level defined by the collision frequency, while for the case with no scattering the losses stop after initial burst and all electrons are mirror trapped. The mean energy of electrons before starting their heating is constant (~50 eV) for the non-scattering case and it rises from 50 to 200 eV for collisional electrons – this artificial heating is caused by higher rate of electron losses for the less energetic electrons.

Moments of switching on the microwave heating are shown in Fig.1 by arrows. Magnitude of the microwave electric field is set to 100 V/cm in these runs. For the no-scattering situation, losses of electrons increase for a short time at the moment of starting the heating due to RF-induced diffusion in velocity space, then losses almost completely disappear. For the collisional electrons, losses start to decrease steadily, following the mean energy increase.

 In a few microseconds after starting the heating, almost all electron trajectories are limited by a surface close to the ECR. A large amount of non-collisional electrons do not cross the ECR surface being trapped inside; these electrons are not heated and remain with their initial energies. Collisional electrons diffuse in space and start at some moment to interact with microwaves at the ECR surface even if their initial trajectories did not cross the resonance layer.

The heating rate (slope of the mean energy time dependence) is largest at the moment of starting the heating and decreases while energies of electrons are increased (Eq.2). For collisional electrons the heating rate decreases slower than for the non-collisional electrons. Time to reach equilibrium both in the mean energy and in life time of electrons is a few milliseconds. Mean electron energy saturates for the collisional electrons at much higher level than for non-collisional electrons.

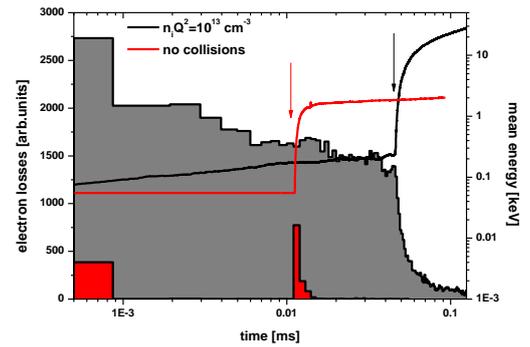

Fig.1. Time dependence of the electron loss rate (left scale, shaded curves) and mean electron energy (right scale) for electrons with and without scattering in collisions with ions.

Steady energy distribution of the collisional electrons at $n_iQ^2=10^{13}$ cm$^{-3}$ and magnitude of electric field of 100 V/cm is shown in Fig.2, combined with the energy distribution for the non-collisional electrons. The distributions are scaled to have the same integral value. The collisional distribution has the bell-like shape with mean energy around 75 keV. There is a narrow peak in EEDF at very low energies consisting of less than 1% of all particles. Without collisions, the distribution is peaked at low energies with a long tail of up to 100 keV. We see strong influence of the collisional scattering on the shape of the energy distribution and on the mean electron energy.

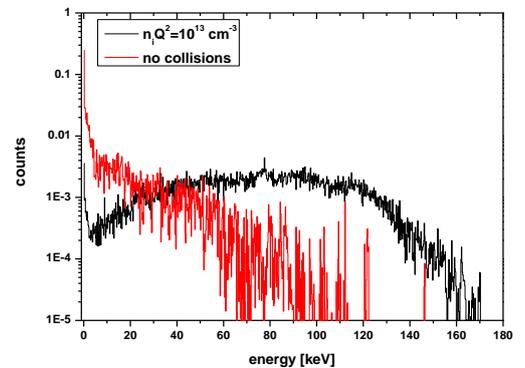

Fig.2. Steady electron energy distribution for electrons with ($n_iQ^2=10^{13}$ cm$^{-3}$) and without collisions.

Describing the electron component in terms of [7], for the collisional spectra most of electrons are *hot* with small contribution of *cold* electrons. The calculated mean energy of EEDF is close to what had been measured as the spectral temperature. The non-collisional spectrum is basically the same as reported in [8].

The collisional electron life time for the situation presented in Fig.2 is 0.1 ms. Most of the lost electrons have relatively low energies and the total losses are governed by losses of cold electrons. The highest probability for electron to be lost is immediately after its creation and while it is heated: scattering rate is highest for the low energetic particles and almost negligible for the run-away electrons. Thus, the electron life time depends on how fast the electrons are heated after their creation.

These considerations are supported by dependencies shown in Fig.3. There, life time of electrons (left scale) and mean energy of the lost electrons (right scale) are plotted as functions of the microwave electric field's magnitude for collisional electrons.

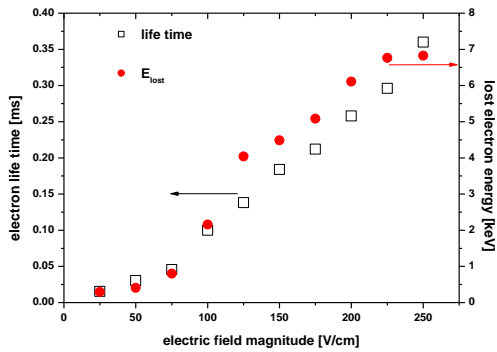

Fig.3. Mean life time of electrons (left scale, open squares) and energy of lost electrons (right scale) as function on magnitude of microwave electric field. The scattering frequency is calculated for the constant factor $n_iQ^2=10^{13}$ cm$^{-3}$.

Increase in the electric field results both in increase of the electron life time and of the mean energy of lost electrons; the dependencies are almost linear. For the lowest shown magnitude of 25 V/cm the life time is ~0.02 ms and the mean energy is 200 eV only. At the same time, mean energy of those electrons that stay inside the plasma is not changing strongly with changing the field magnitude – it is around 65 keV for $E_0=25$ V/cm and increases up to 72 keV for $E_0=200$ V/cm. There is a large difference between mean energy of electrons in the plasma and mean energy of the lost electrons – the values differ by almost two orders of magnitude.

The calculated electron life time is inversely proportional to the collision frequency, $\tau_e\sim1/(n_iQ^2)$. With lowering the collision frequency to $n_iQ^2=10^{12}$ cm$^{-3}$ and for $E_0=100$ V/cm, mean energy of electrons in the plasma remains constant, energy of the lost electrons increases up to 5 keV compared to 2.5 keV for the larger ion density, and the life time increases to 1 ms, order of magnitude larger than the value in Fig.3.

For such energies of electrons in plasma, relativistic shift in the resonant frequency is not negligible. The mean resonance value of the magnetic field is calculated to be 0.67 T for the mean energies of around 70 keV, exceeding the non-relativistic value of 0.643 T by 5%.

After obtaining these general tendencies of the electron dynamics in ECRIS, we switch to calculations with more realistic spatially resolved ion densities. The ionic part of NAM-ECRIS is run assuming the Maxwell-Boltzmann electron energy distribution, and the electron temperature is selected such that the mean energy corresponds to the values mentioned above, namely in the range of (40-60) keV. The potential dip is located at the magnetic field that corresponds to the relativistically shifted resonance value, 0.67 T.

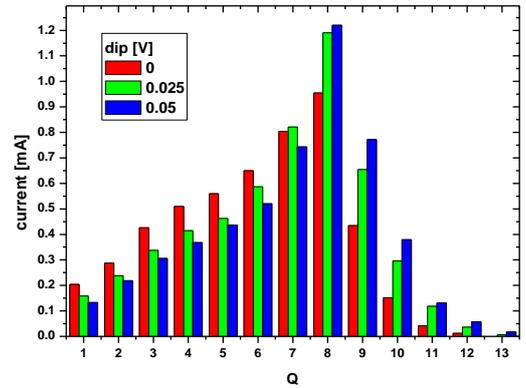

Fig.4. Charge state distribution of the extracted argon ion currents for different potential dip values. Statistical weight of ions is $8\times10^8$, electron temperature is 45 keV.

Calculations are done with fixing the statistical weight of numerical particles and changing the potential dip value. The larger is the potential dip, the larger are the ion life time in plasma, mean ion charge state and electron density. Changes in the extracted ion currents with increasing the potential dip are shown in Fig.4. Here, the statistical weight of the particles is set to $8\times10^8$, electron temperature is 45 keV, and the dip is changing from 0 to 0.05 V. Current of Ar$^{8+}$ ions is around 1 mA in these conditions and saturates at large dip values, while the higher charge state currents are increasing steadily.

After each run with fixed potential dip value, ionic densities are imported into the electron part of the code and calculations are done with varying the electric field magnitude such as to obtain the electron life time close (±25%) to the calculated ion life time. Mean energy of electrons in the plasma is calculated and the ion part is run again if the energy differs from the value that corresponds to the initially selected electron temperature. This process is repeated iteratively if needed; in practice we observe very small variations in plasma parameters when changing the electron temperature in the above-mentioned range – ionization rates and ion heating rate change slowly for such high electron temperatures.

Basically, in the described way we obtain correlation between the plasmas with given global ion life time and magnitude of microwave electric field (microwave power injected into the source) needed to ensure that the global electron life time is at the same level. Reversing the sequence, we observe changes in the ion life time and corresponding changes in global plasma parameters with changing the microwave electric field at resonance, which is proportional to the injected microwave power.

Plasma with potential dip value equal to zero defines the lowest life time for ions that can be observed basing on the potential dip regulation of the ion losses. To obtain plasmas with lower ion life time, we need in a mechanism for degrading the ion confinement. For these cases, we introduce into calculations the electric field inside the plasma, which is directed such as to expel ions toward the source walls. Our approximation is that the electric field is pointing in the direction of increasing magnetic field and is constant all over the plasma volume. The field value is input for the ionic part of the code; it is varied in the range of up to 0.01 V/cm. The larger is the electric field, the lower are the ion confinement times. Confinement of the highly charged ions degrades faster with increasing the field compared to the lowly charged ions.

Results of this iterative procedure are presented in Fig.5, where dependence of the ion/electron life time on the magnitude of microwave electric field is shown.

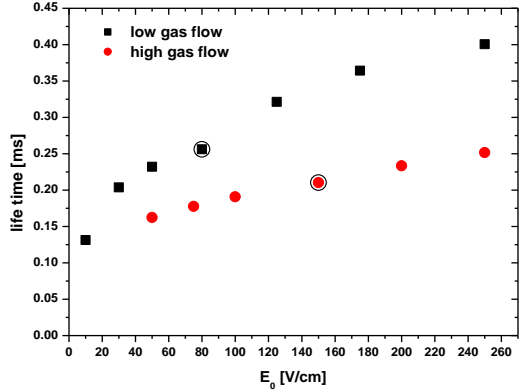

Fig.5. Electron/ion life time as function of the microwave electric field magnitude $E_0$ for argon plasma parameters obtained with different gas flows into the source (low and high gas flows correspond to statistical weights $8\times10^8$ and $12\times10^8$).

The life times are plotted for two statistical weights of ions, $8\times10^8$ (low gas flow, black squares) and $12\times10^8$ (high gas flow, red circles). The highlighted points correspond to the plasmas with zero potential dip value. Points with the larger life times are obtained with the potential dips incremented with step of 0.025 V. Points with the lower life times are obtained by varying the expelling electric field inside the plasma.

Gas flows into the source differ for the chosen statistical weights by 50%, roughly proportional to the weight ratio: 1.3 particle-mA for the weight of $8\times10^8$ and 2.0 particle-mA for $12\times10^8$. Mean electron density inside the ECR zone changes with increasing the weight from $9\times10^{11}$ cm$^{-3}$ to $1.2\times10^{12}$ cm$^{-3}$. Larger electron and ion densities result in decrease of the ion life time due to increased rate of ions heating in electron-ion collisions. Larger magnitudes of the microwave electric field increase the electron life time, which results either in decrease of electric field inside the plasma that pushes ions out the plasma or in increase of the potential dip that retards the ions, depending on the mode of ion confinement. Electron temperature increased with the electric field magnitude from 45 keV to 52 keV in the studied range of the magnitude's variation.

The values of electron life time were given for total electron losses out of the plasma. If focusing on the losses of energetic electrons, much larger life times are observed: for electrons with energies above 10 keV, life time is 10 ms for the electric field magnitude $E_0$= 80 V/cm at the weight of $8\times10^8$ and dip of 0 V and it decreases to 4 ms for $E_0$=250 V/cm and dip of 0.075 V.

Extracted currents of $Ar^{8+}$ and $Ar^{9+}$ ions are shown as a function of the microwave electric field in Fig.6. Currents are plotted for the same data sets as in Fig.5. For electric field magnitudes above ~100 V/cm, currents of ions for both shown gas flows tend to saturate, with the larger currents for the larger gas flow. For relatively small field magnitudes, increase in the gas flow does not result in increase for the $Ar^{8+}$ ion current, for the higher charge states the currents decrease. This is consistent with the experimental observations concerning reaction of the highly charged ion currents on the injected microwave power and gas flow into the source.

In calculations of the electron dynamics we derive the mean energy of electrons that are lost from the plasma. By combining this value with the calculated electron fluxes out of the plasma, we obtain the power associated with electron losses out of the plasma. The dependencies of the power on the microwave electric field are shown in Fig.7 for the same data as in Figs.5-6.

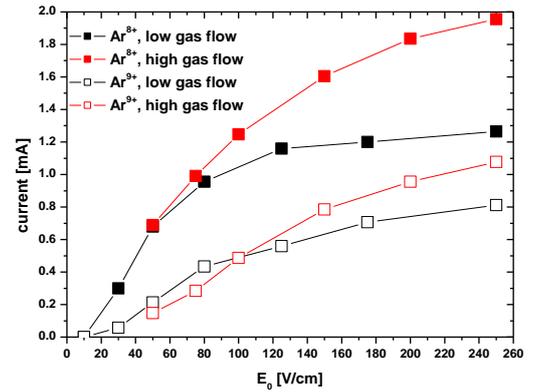

Fig.6. Extracted currents of $Ar^{8+}$ and $Ar^{9+}$ ions as function of the microwave electric field magnitude $E_0$ for argon plasma parameters obtained with different gas flows into the source.

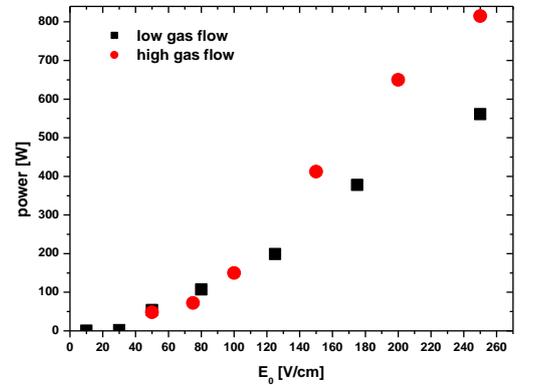

Fig.7. The power carried away by lost electrons as a function of the microwave electric field magnitude for argon plasma obtained with different gas flows into the source (statistical weights $8\times10^8$ and $12\times10^8$).

Increase in the power losses out of the plasma is seen when increasing the microwave field magnitude. The power is around 100 W for the electric field of 100 V/cm and reaches the values close to 1 kW for the larger fields. The power increases with the higher gas flow and larger electron current toward the walls. For the same microwave electric field of 250 V/cm, the power carried away by the lost electron is larger by 50% for the 2 pmA of the injected argon flux compared to the case of the lower flux.

Basically, we see that to reach ~2 mA of $Ar^{8+}$ extracted ion current, we need to increase the gas flow by 50% comparing to the flow that provides saturation at the level of 1 mA. At the same time, the electric field magnitude should be doubled to reach these ion currents, which means that the injected microwave power should be increased by factor of four. In practice, around 1 kW of the microwave power should be injected into DECRIS-SC2 18 GHz source to extract 1 mA of $Ar^{8+}$ ion current. Then we can estimate that more than 5 kW of the power is required to reach the level of ~2 mA of the $Ar^{8+}$ current, taking into account the increased channel of power losses due to the lost electrons.

The described features of electron dynamics allow explaining the so-called afterglow effect [21] in ECRIS. Experimentally, it is observed that short (~1-5 ms) pulse of highly charged ion currents that sufficiently exceeds the DC currents of ions can be produced after switching the microwave heating off. We see that the electron losses in ECRIS are mainly determined by fluxes of new-born *cold* electrons before they are heated to high energies. Switching off the heating will result in increase of electron fluxes out of the source plasma followed by the corresponding loss of ion confinement and burst in the extracted ion currents. The process is illustrated by Fig.8, where the total electron flux out of plasma is plotted as a function of time with and without the electron heating. The dependence is obtained for the statistical weight of ions equal to $8\times10^8$ and for the electric field magnitude of 175 V/cm for the argon plasma with the potential dip of 0.05 V. Moment of switching the heating off is labeled with arrow.

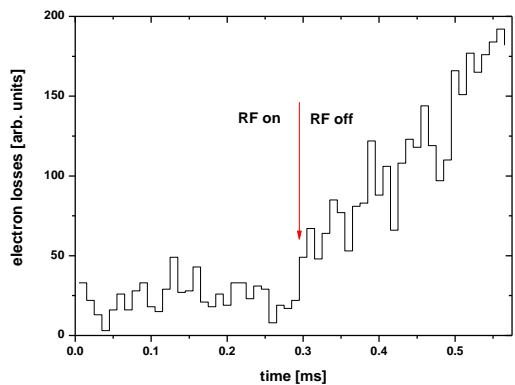

Fig.8. Electron losses out of the plasma before and after switching the microwave heating off.

The electron losses increase steadily after terminating the microwave heating; time to double the loss rate is around 0.1 ms. Completely different picture is seen in calculations if counting only for energetic electron losses – electron life time for electrons with energies larger than 10 keV is ~7 ms with the RF heating and it is increased up to 28 ms after switching the heating off. This indicates that the energetic electrons are lost mainly because of the RF-induced diffusion in velocity space; termination of these losses in afterglow stage will manifest itself as a drop in X-ray emission out of the source, which is observed in experiments [6]. Calculations are done without taking into consideration a decrease in plasma density after switching the electron heating off. In more realistic situation, electron losses will start to decrease at some moment due to decay of the plasma.

Experimentally, it was observed that output currents of ECRIS can be increased if heating is done with using microwaves with two different frequencies. Mixing 18 GHz and 14 GHz microwaves in the discharge results in noticeable gains in the range of (20%-50%) for the currents of the highly charged xenon ions even if power of 14 GHz microwaves is less than 10% of the total injected power [22].

Coming back to analysis of data in Fig.1, we notice that the heating rate for collisional cold ($\gamma=1$) electrons at the moment of starting the heating is 0.86 keV/μs for $B_{res}=0.643$ T. These calculations are repeated for $B_{res}=0.5$ T without changing the magnetic field profile and for the same magnitude of electric field of 100 V/cm. The electron velocity kick should be scaled by 13% for 14 GHz microwaves according to Eq.2. The important factor is that the ECR volume is much smaller for $B_{res}=0.5$ T compared to the $B_{res}=0.643$ T – length of the ECR zone at the source axis is 3 cm for 14 GHz and 7.5 cm for 18 GHz microwaves.

The heating rate for the cold electrons is calculated to be 2.4 keV/μs for 14 GHz, much larger than for the default 18 GHz. The mean electron energy after reaching saturation is also noticeably exceeding (by the factor of ~40%) the calculated energy for $B_{res}=0.643$ T, being at the level of 95 keV. For the given collision frequency of electrons, the faster heating rates mean that the life times for electrons in plasma are larger.

Reason for the boost in heating rate is the smaller length of ECR volume, which results in higher frequency of kicking the electrons while they bounce between their turning points located close to the resonance surface. The negative manifestation of the smaller ECR volume is that the plasma volume also becomes be smaller, with lower ion contents for the fixed electron density. The same approach is applicable when analyzing the source response to changes in the ECR volume by changing the magnetic field profile: smaller sizes of the volume assume faster electron heating rates (longer electron life times) counteracted with smaller amount of ions available for extraction.

Currents of the extracted argon ions were calculated for the $B_{res}=0.54$ T (relativistic value of the resonant field for $\gamma=1.08$) and the potential dip of 0 V for the same statistical weight of $8\times10^8$ as in Fig.4. Electron temperature was set to 62 keV following the electron dynamics simulations. Current of $Ar^{8+}$ ions is 0.48 mA for these settings, much smaller than 0.96 mA for the corresponding case with 18-GHz heating.

Compromise between increasing the electron heating rate and shrinking the ECR volume can be found if intensity of 14 GHz microwaves is small compared to the main waves. We calculate that if power of 14-GHz waves is 25% of the total injected microwave power, the cold electron heating rate is 1.6 keV/μs, smaller than for heating with the single 14-GHz waves, but still twice larger of the value for the single 18-GHz heating. At the same time, electrons occupy the same volume as for 18-GHz waves and the plasma volume is not decreasing. Simulations show that the potential dip value for argon plasma should be 0.01 V for the mix of two frequencies in combination ($E_0$(18 GHz)=70 V/cm + $E_0$(14 GHz)=40 V/cm). This combination corresponds to the same total microwave power as for $E_0$=80 V/cm in Fig.4. At this, currents of argon ions with the charge state above 8+ are increased by (10-50)% compared to single 18-GHz heating, being especially boosted for the highest charge states.